\documentclass[aps,pre,twocolumn,groupedaddress,showpacs,floatfix]{revtex4}
\newcommand{\Na}{${\textrm {Na}}^+$}
\newcommand{\K}{${\textrm {K}}^+$}

\newcommand{\Cl}{${\textrm {Cl}}^-$}
\DeclareMathVersion{euler}
\SetSymbolFont{letters}{euler}{U}{eur}{m}{n}
\newcommand{\micro}{\mathversion{euler}$\mu$\mathversion{normal}}
\newcommand{\um}{{\micro}m}
\newcommand{\us}{{\micro}s}
\newcommand{\uF}{{\micro}F}
\newcommand{\uA}{{\micro}A}

\newcommand{\uFcmsq}{\uF/$\textrm{cm}^2$}

\newcommand{\mScmsq}{{mS}/$\textrm{cm}^2$}

\newcommand{\Ohmcm}{$\Omega \cdot \textrm{cm}$}
\newcommand{\GNabar}{\bar{G}_\textrm{{\tiny Na}}}

\newcommand{\GKbar}{\bar{G}_\textrm{{\tiny K}}}
\newcommand{\GNa}{G_\textrm{{\tiny Na}}}
\newcommand{\GK}{G_\textrm{{\tiny K}}}
\newcommand{\GL}{G_\textit{{\tiny L}}}
\newcommand{\GLNa}{G_{\textit{{\tiny L}}\textrm{{\tiny Na}}}}
\newcommand{\GLK}{G_{\textit{{\tiny L}}\textrm{{\tiny K}}}}

\newcommand{\ENa}{E_\textrm{{\tiny Na}}}
\newcommand{\EK}{E_\textrm{{\tiny K}}}
\newcommand{\ECl}{E_\textrm{{\tiny Cl}}}
\newcommand{\EL}{E_\textit{{\tiny L}}}
\newcommand{\Er}{E_r}

\newcommand{\vAP}{v_\textrm{{\tiny{\it AP}}}}
\newcommand{\Ra}{R_a}

\usepackage[dvips]{graphics}
\begin{document}
\title{Optimization of the leak conductance in the squid giant axon}
\author{Jeffrey Seely}
\author{Patrick Crotty}
\email{pcrotty@colgate.edu}
\thanks{corresponding author.}
\affiliation{Department of Physics and Astronomy, Colgate University, 13 Oak Drive, Hamilton, NY, USA 13346}
\date{\today}


\begin{abstract}
We report on a theoretical study showing that the leak conductance density,
$\GL$, in the squid giant axon appears to be optimal for the 
action potential firing frequency.  More precisely, the standard assumption that the
leak current is composed of chloride ions leads to the result that the experimental 
value for $\GL$ is very close to the optimal value in the Hodgkin-Huxley model 
which minimizes the absolute 
refractory period of the action potential, thereby maximizing the maximum firing 
frequency under stimulation by sharp, brief input current spikes to one end of
the axon.  The measured value of $\GL$ also appears to be close to optimal for the 
frequency of repetitive firing caused by a constant current input to one end of the 
axon, especially when temperature variations are taken into account.  If, by contrast,
the leak current is assumed to be composed of separate voltage-independent sodium and 
potassium currents, then these optimizations are not observed.  
\end{abstract}

\pacs{87.19.lb,87.19.ll,87.19.lo,02.60.Pn}
\maketitle

\section{Introduction}

A considerable amount of evidence has emerged in recent years to show that many of 
the parameters which govern the structure and function of biological nervous 
systems are at optimal values for metabolic energy consumption, information 
rates, or some combination thereof, presumably because of evolutionary 
pressures \cite{levy_1996, levy_2002, goldberg_2003}.  Many of these studies have 
focused on the squid giant axon because of its well-known and relatively simple 
properties.  Hodgkin
and Adrian hypothesized as early as the 1970s that the channel densities in the squid
giant axon are at values that maximize the action potential velocity
\cite{adrian_1975, hodgkin_1975}, although
more updated axon models have called this into question \cite{sangrey_2004} and 
suggested that axons are optimized for the energy associated with the action potential
instead \cite{crotty_2006b, levy_2006}.  There are a large number of independently 
variable parameters that significantly affect the functioning of the squid giant 
axon, only a few of which have been systematically investigated for possible
optimizations.  Here, we present results for one of them:  the leak conductance.

The voltage-independent leak conductance is one of three conductances known 
to be present in the squid giant axon.  With a measured value of
about 0.3 \mScmsq, it is much smaller in magnitude than the maximum voltage-gated 
\Na{} (120 \mScmsq) and \K{} (36 \mScmsq) conductances, yet it nevertheless 
plays an important role in the electrical stability of the axon.  Leak
conductances are known to be present in many other kinds of neurons as well, such as
molluscan pacemaker cells \cite{hille_2001}.

Because of its small size, there has long been debate about the exact nature 
of the leak conductance, with some of the debate centering on how much of it is
due to voltage-gated \K{} and \Na{} channels that remain open at rest 
\cite{chang_1986, clay_1988}.
However, a nonselective, voltage-independent cation channel protein has recently been
conclusively identified in mammalian neurons \cite{lu_2007}.   In squid giant
axons, the closeness of the leak reversal potential to the equilibrium potential of 
\Cl{} has traditionally been taken to indicate that chloride ions are a significant
contributor to the current, although there may be others \cite{hodgkin_1952b}.

Given the possibility that the leak conductance is an easily evolvable
parameter capable of influencing the electrical 
properties of axons, it is natural to ask whether the leak conductance is 
at an optimal value for some quantity related to information processing or energy 
consumption.  Here, we present results showing that the leak conductance is
near-optimal for the absolute and repetitive firing frequency of the axon if the
leak current is assumed to be chloride, but not if it is sodium/potassium.
For completeness we also investigated the effects of $\GL$ on the relative 
refractory period, which is substantially harder to calculate numerically.  
We did not find strong evidence that $\GL$ is optimal in this case for either 
channel model.


\section{Methods}

\subsection{The Hodgkin-Huxley Model}

The squid giant axon, about 0.5 mm in diameter, is one of the largest
axons in nature.  It innervates muscles in the squid mantle, and its action
potentials cause the muscles to contract, expelling a brief jet of water
and allowing the squid to move away quickly from danger.  The axon is postsynaptic
to neurons in the dorsal magnocellular lobe, and it is indirectly connected to
the ventral magnocellular lobe, which integrates sensory input.  
  
Fortunately, the squid giant axon is also one of the simplest known axons, being 
unmyelinated and having only two voltage-gated ion channels 
with relatively straightforward kinetics.  As such, we
arguably know more about the squid giant axon than any other neural
system, and it is possible to model it to a high degree of biological
accuracy.  The Hodgkin-Huxley (HH) model, based on the experiments
of A. L. Hodgkin and A. F. Huxley in 1952 \cite{hodgkin_1952}, remains very useful, 
although subsequent refinements to the sodium and potassium channel kinetics have been
made \cite{vandenberg_1991, clay_2005}.  In our study, we used the traditional 
version of the HH model described below.

The model treats the squid giant axon as a cylinder of length $L$ and uniform diameter
$d$.  For most of our simulations, $L$ was set to 0.8 cm.  The diameter
was generally in the range
of 300-600 \um.  (These dimensions are typical of the biological ones.)  The 
cylinder has an axial resistivity $\Ra$ representing
the axoplasm.  This was generally set to 35.4 \Ohmcm{} after Hodgkin and 
Huxley's measurement.  Transmembrane
currents flow through two voltage-dependent \Na{} and \K{} conductances, 
$\GNa$ and $\GK$, and a voltage-independent leak conductance, $\GL$.
The Hodgkin-Huxley experimental measurements of $\GL$ ranged from 0.13 to
0.5 \mScmsq, with an average of about 0.26 \mScmsq \cite{hodgkin_1952b}.  It
was primarily this value that we varied in our simulations.  The 
voltage-dependent conductances are functions of time and, indirectly, the 
membrane potential:
\begin{equation}
\label{eq_GNa}
\GNa = \GNabar \left[m(t)\right]^3 h(t) \,,
\end{equation}  
\begin{equation}
\label{eq_GK}
\GK = \GKbar \left[n(t)\right]^4 \,,
\end{equation}

\noindent where $m$ and $h$ are state variables
representing the fraction of \Na{} channel subunits in the open
and non-inactivated states, respectively, and $n$ is the state variable 
representing the fraction of open \K{} channel subunits (squid \K{} 
channels do not inactivate).  Each
of the four subunits in a channel has to be open or non-inactivated
in order for the channel to pass ions.  Thus, $m^3h$ and $n^4$ are, 
respectively, the fraction of \Na{}
and \K{} channels which are open.  These state variables evolve
according to equation \ref{eq_state_ode} below.  The maximal conductances,
$\GNabar$ and $\GKbar$, obtain when all the channels are open.  We used
Hodgkin and Huxley's experimental values of $\GNabar = 120 \textrm{
\mScmsq}$ and $\GKbar = 36 ~ \textrm{\mScmsq}$.

The cell membrane has a constant intrinsic capacitance of approximately
$C_0 = 0.88 \, \text{\uFcmsq}$ \cite{gentet_2000}.  The 
voltage-gated sodium channels also contribute a phenomenological capacitance, 
$C_g = 0.13 \, \text{\uFcmsq}$, the so-called ``gating'' 
capacitance \cite{hodgkin_1975}.  (In principle, the 
voltage-gated potassium channels contribute a gating capacitance too, but 
this is so much smaller than the sodium gating capacitance that it can be
neglected.)  

Electrical excitations of the axon are described by
four coupled differential equations.  The first of these is a modified
version of the cable equation:
\begin{eqnarray}
\label{eq_HH_pde}
\frac{d}{4 R_a} \frac{d^2 V}{dx^2} = \left(C_0 + C_g\right) \frac{dV}{dt}
+ \GNabar m^3 h \left(V - \ENa \right)  \nonumber \\
+ \GKbar n^4 \left( V - \EK \right)
+ \GL \left( V - \EL \right)
\end{eqnarray}

\noindent where $V$ is the cross-membrane potential, with the
extracellular side taken as ground.  The \Na{} and \K{} reversal
potentials, $\ENa$ and $\EK$, are determined by the ionic concentration
gradients across the membrane:  the values we used, $\ENa$ = 50 mV and 
$\EK$ = -77 mV, are typical of the squid giant axon.  The leak reversal
potential, $\EL$, is determined by the ion(s) which pass through the
leak channels and is experimentally around -55 mV \cite{hodgkin_1952b}.  
We discuss our leak channel assumptions in greater detail below.

The other three differential equations govern the gating variables:
\begin{equation}
\label{eq_state_ode}
\frac{ds}{dt} = \alpha_s(V) \times (1 - s) - \beta_s(V) \times s \,,
\end{equation}

\noindent where $s$ = $n$, $m$, or $h$.  The rate coefficients,
$\alpha_s$ and $\beta_s$, are the fraction of $s$ subunits per unit time
switching from closed/inactivated to open and open to closed/inactivated,
respectively.  These rates were empirically measured by Hodgkin and Huxley
as:
\begin{eqnarray}
\label{eq_alpha_m}
\alpha_m\left(V\right) = \phi \times 0.1 \times \frac{-\left(V + 40\right)}{
\left( e^{-\left( V + 40 \right) / 10} - 1 \right)}
~~~ \left[\textrm{ms}^{-1}\right]
\\
\label{eq_beta_m}
\beta_m\left(V\right) = \phi \times 0.4 \times e^{-\left(V + 65\right)/18} 
~~~ \left[\textrm{ms}^{-1}\right]
\\
\label{eq_alpha_h}
\alpha_h\left(V\right) = \phi \times 0.07 \times e^{-\left(V+65\right)/20} 
~~~ \left[\textrm{ms}^{-1}\right]
\\
\label{eq_beta_h}
\beta_h\left(V\right) = \phi \times \frac{1}{\left( e^{-\left(V+35\right)/10} 
+ 1 \right)} ~~~ \left[\textrm{ms}^{-1}\right]
\\
\label{eq_alpha_n}
\alpha_n\left(V\right) = \phi \times 0.01 \times \frac{-\left(V + 55
\right)}{\left( e^{-\left(V + 55\right) / 10} - 1 \right)} ~~~ \left[\textrm{ms}^{-1}\right]
\\
\label{eq_beta_n}
\beta_n(V) = \phi \times 0.125 \times e^{-\left( V + 65 \right) / 80}
~~~ \left[\textrm{ms}^{-1}\right]
\end{eqnarray}

\noindent with $V$ in mV.  The temperature coefficient $\phi$ is
\begin{equation}
\label{eq_phi}
\phi = 3^{\left( T - 6.3 \right)/10}
\end{equation}

\noindent with $T$ in $^\circ$C.  It should be noted that the openings and 
closings of individual ion channels are stochastic in nature; Eq. 
(\ref{eq_state_ode}) describes the average behavior of a large ensemble of 
$s$-subunits.

Eqs. (\ref{eq_HH_pde}) and (\ref{eq_state_ode}) are highly nonlinear
(Eq. (4) because of the forms of the rate coefficients (5)-(10)); and, without 
significant approximations, they are analytically intractable.
However, it is known that they have a unique solution describing
a single voltage spike, or ``action potential,'' propagating at a uniform
velocity $\vAP$ along the axon.  The action potential velocity is a function of the
different biophysical parameters in Eqs. (\ref{eq_HH_pde}-\ref{eq_beta_n}),
though it does not have an exact analytical form and must generally
be either approximated or determined numerically.

Our simulated axon contained 1000 isopotential segments, each of length
100 \um.  Equations \ref{eq_HH_pde} and \ref{eq_state_ode} 
(for $n$, $m$, and $h$) were solved simultaneously in each segment
using an implicit backward Euler method.  Our time step was 1 \us.
We verified that the time and spatial resolutions were sufficiently fine so as to
not significantly influence our results.  We assumed $T = 18.5 ~  ^\circ\textrm{C}$ 
unless stated otherwise.

\subsection{The Leak Channel}

We tested two different assumptions about the nature of the leak channel.  
We assumed first that it is a voltage-independent \Cl{} conductance, in which case 
the leak reversal potential, $\EL$, is just the equilibrium potential of chloride: 
\begin{equation}
\label{eq_EL_Cl}
\EL = \ECl \approx -55 ~\textrm{mV} ~ ;
\end{equation}

\noindent We identify this case as ``\Cl{} leak'' in the figures.  In our simulations,
we typically varied the value of  $\GL$ while keeping all the other parameter 
values the same.  This has the effect of changing the resting potential of the axon 
(see Figure \ref{fig_HH_Cl__GL_Er}) and, indirectly, the maximum frequency at which
action potentials can fire.  We numerically calculated the new resting potential 
and set the axon to this value at the beginning of our simulations before action 
potentials were evoked.

In the second case, we assumed that the leak channel consists of two 
voltage-independent \Na{} and \K{} conductances, in which case $\EL$ is determined 
by a weighted average of $\ENa$ and $\EK$:
\begin{equation}
\label{eq_EL_KNa}
\EL = \frac{\GLK \EK + \GLNa \ENa}{\GLK + \GLNa} ~ .
\end{equation}

\noindent In this case, which is identified as ``\Na/\K{} leak'' in the figures, the 
total leak conductance is just the sum of the sodium and potassium leak conductances:
\begin{equation}
\label{eq_GL_KNa}
\GL = \GLNa + \GLK ~ .
\end{equation}

Since only $\GL$ and $\EL$ appear in the equation of motion for $V$, Eq. 
(\ref{eq_HH_pde}), the refractory periods are not affected by 
whether it is \Cl{} or \Na/\K{} that goes through the leak channel if only $\GL$ is 
varied.  That is, any optimization results involving the leak conductance alone that 
obtain for \Cl{} leak channels should obtain for \Na/\K{} leak channels as well -- 
or, in fact, for a leak current composed of any combination of permeant ions, 
provided that their overall reversal potential is approximately $\ECl$.  In order to 
more completely distinguish between the two cases, we added the further requirement 
in the \Na/\K{} leak channel case that whenever the value of $\GL$ was altered, the 
value of $\EL$ was altered as well so as to keep the overall resting potential, 
$\Er$, at -65 mV.  This is equivalent to altering the ratio of the \Na{} leak
conductance, $\GLNa$, to the \K{} leak conductance, $\GLK$, while keeping the 
reversal potentials $\ENa$ and $\EK$ the same.

Thus, the mathematical distinction between the two models is that for \Cl{} leak
channels, only $\GL$ is varied in Eq. (\ref{eq_HH_pde}), which in turn causes
the resting potential to vary as shown in Fig. (\ref{fig_HH_Cl__GL_Er}).  
For \Na/\K{} channels, by constrast, both $\GL$ and $\EL$ are varied in tandem so
as to keep the resting potential at -65 mV.  

The next generalization of these models would be to allow both $\GL$ and $\EL$ to 
vary independently, which would also change $\Er$ to varying degrees:  for example, 
if $\GL$ were very large, $\Er$ would be pulled towards $\EL$, while if 
$\GL$ were very small, the value of $\EL$ would have little effect on $\Er$.
Systematic investigations of the optimzations discussed here in such an extended 
model would be challenging due to the high computational costs of multidimensional 
parameter sweeps.

\subsection{Simulations}

All of our simulations were done using the NEURON/NMODL neuronal modeling
language \cite{hines_1997} and auxiliary parameter-sweeping codes written in 
C and Python.  We systematically varied $\GL$ in Eq. (\ref{eq_HH_pde}) and 
occasionally other parameters, which are described in further detail in Section 
III, in order to determine how they influence the absolute and relative refractory 
periods of the action potential and the frequency of repetitive firing.  

Simulated action potentials were evoked in two different ways.  When studying
the absolute and relative refractory periods, action potentials were evoked by
1 A, 1 \us{} duration current injections into one end of the axon; we verified that 
these values 
were sufficiently brief and large so as to not influence the refractory periods.
They are referred to throughout the text as ``current spike-evoked'' action 
potentials.  The absolute refractory period, $T_{abs}$, was determined by finding 
the maximum time between successive current injections such that only one action 
potential resulted.  We verified that for inter-injection times just above the 
absolute refractory period, two action potentials were produced and both propagated 
down the full length of the axon, as expected.  The maximum possible action
potential firing frequency, i.e., the maximum frequency at which the axon can
be driven with current spike inputs, is then the reciprocal of the absolute refractory 
period,
$f_{max}$:
\begin{equation}
\label{eq_Tabs_fmax}
f_{max} = \frac{1}{T_{abs}}
\end{equation}

If the time interval between the input current spikes is greater than $T_{abs}$
but less than a certain value $T_{rel}$, called the relative refractory period, then 
while a second action potential is generated, it is generated in the wake
of the first one, when the membrane and the voltage-gated ion channels have not
yet returned to their resting states.  The result is that the two action
potentials interfere with each other:  the velocity of the second is different
from that of the first, and the distance between the action potentials -- or,
equivalently, the time between their peaks as measured at a single point along the
axon -- changes as they move down the axon.  Information encoded in the 
distribution of intervals between action potentials can therefore be corrupted if 
they are too close together.
If $T_i$ is the time between the two current spikes and $T_{AP}$ is the time between 
the peaks of the two resulting action potentials as measured at some point further 
down the axon, then we can define the ``interval shift'' as 
$\Delta T \equiv T_{AP} - T_{i}$.  For $T_i \ge T_{rel}$, $T_{AP} = T_i$ and so
$\Delta T = 0$.

It is somewhat difficult to determine $T_{rel}$ numerically.  Action potentials in
the Hodgkin-Huxely model have a phase of small, damped oscillations around the
resting potential after the peak, which means that a second closely-following
action potential can be either sped up or slowed down depending on which
part of an oscillation it falls into.  As a result, the interval shift $\Delta T$
itself oscillates around 0 as a function of $T_i$ (Fig. \ref{fig_rrel_difficulties}
and \cite{crotty_2006}), and this
causes $T_{rel}$ as a function of $\GL$ to have a jagged, discontinuous appearance
(Fig. \ref{fig_HH__x_0p8__dt_0p2__GL_frel}).  Moreover, jitter noise in
biological axons (caused by phenomena such as ion channel flicker) puts a nonzero
lower bound on the timing resolution of consecutive action potentials.  

Thus, it is both computationally easier and probably more biologically relevant to
define a relative refractory period as a function of the maximum interval shift:
for all values of $T_i$ greater than $T_{rel} \left( \Delta T_{max} \right)$, by
definition, $\left| \Delta T \right| = \left|T_{AP} - T_i\right| \le \Delta T_{max}$,
where $\Delta T_{max} > 0$.  The reciprocal of $T_{rel} \left( \Delta T_{max} \right)$ 
gives the maximum frequency, $f_{rel} \left( \Delta T_{max} \right)$, at which the axon 
can be driven with an interval shift no larger than $\Delta T_{max}$:
\begin{equation}
\label{eq_Trel_frel}
f_{rel}\left( \Delta T_{max} \right) = \frac{1}{T_{rel}\left( \Delta T_{max} \right)}
\end{equation}

\noindent We determined $T_{rel} \left( \Delta T_{max} \right)$ for values of 
$\Delta T_{max}$ ranging from 1 \us{} (the numerical resolution of our simulations)
up to 1 ms.

To determine the repetitive firing frequency, $f_r$, we simulated a constant
(time-independent) current input, $I_{DC}$, to one end of the axon.  We then
tested whether regular, repetitive firing resulted and, if so, 
measured the time between equivalent points on successive action potentials 20 ms 
after the beginning of the input current, by which time any initial transients
had long since disappeared.  The reciprocal of this time was then $f_r$, which in 
general depended on the value of $I_{DC}$ as well as that of $\GL$.  We refer to 
$f_r$ as the ``repetitive firing frequency.''

We generally measured the time intervals between successive action potentials at
a point 8 cm down the axon from the input stimuli.  We verified that our optimization
results were insensitive to the actual location of this point as long as it
was outside a small region near the current injection site.


\section{Results}

\subsection{Resting Potential and Individual Action Potentials}

\begin{figure}
\resizebox{3.25in}{2.5in}{\includegraphics{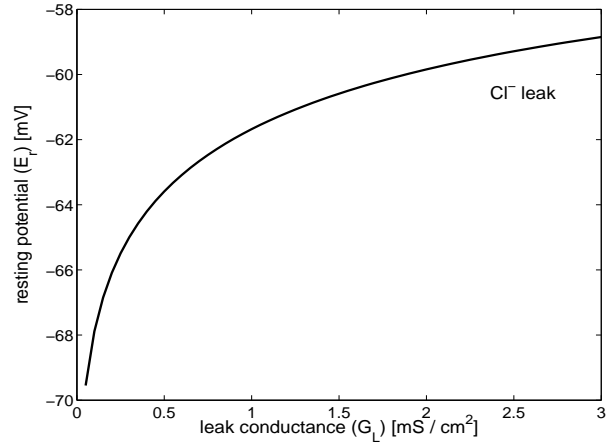}}
\caption{\label{fig_HH_Cl__GL_Er} Despite its small magnitude, $\GL$ has a
significant effect on the resting potential of the axon, $\Er$, defined as the 
potential at which there is no net transmembrane current.  Here we plot $\Er$ as
a function of $\GL$ while keeping all other parameters fixed (this therefore 
corresponds to the case of \Cl{} leak channels.)}
\end{figure}

In addition to the effects of $\GL$ on firing frequencies, which is the main
focus of this study, $\GL$ also affects the shapes of individual action potentials
and, in the case of \Cl{} channels, the overall resting potential of the axon.
The resting potential $\Er$ is the voltage at which the sum of all the 
steady-state ionic currents is 0 (Fig. \ref{fig_HH_Cl__GL_Er}).  It is
therefore determined by the leak current as well as the small currents through the
voltage-gated sodium and potassium channels, which are almost (but not entirely)
closed at $\Er$.  Physically, the resting potential is determined by the
channel densities, the ionic concentration gradients, and the steady-state
conformational configurations of the voltage-gated channels.

Using the standard (and reasonably well-established) values of these other 
parameters, as $\GL$ increases from 0.05 to 3 \mScmsq, the resting potential 
increases (in the sense of getting less negative) by roughly 10 mV, with the 
sharpest rate of increase in the range below 
1 \mScmsq.  For \Na/\K{} channels, there is no such dependence of $\Er$ on $\GL$
because, as discussed above, we simultaneously varied the ratio of \Na{} and \K{}
leak conductances in order to keep $\Er$ at -65 mV.

\begin{figure}
\resizebox{3.25in}{2.5in}{\includegraphics{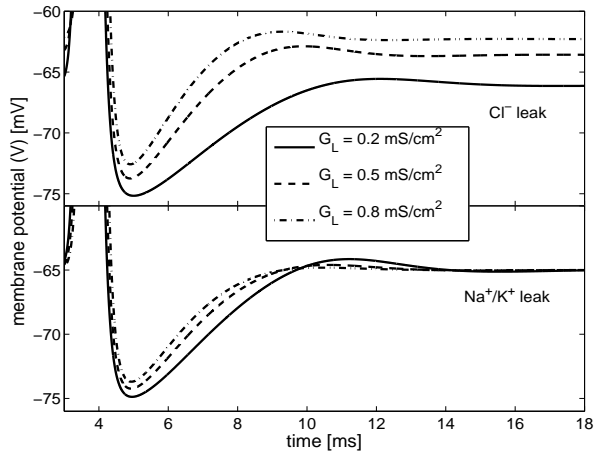}}
\caption{\label{fig_HH_Cl__GL_range} At lower values of the leak conductance,
the axon takes longer to return to rest after a current spike-evoked action 
potential.  Note that the resting potential is always made to be -65 mV for the
\Na/\K{} case.}
\end{figure}

The effects of $\GL$ on individual (current spike-evoked) action potentials 
differ between the two models.  With \Cl{} channels, higher values of $\GL$ lead
to smaller and narrower spikes that have a more pronounced post-peak oscillation.
With \Na/\K{} channels, by contrast, the heights and widths of the main action
potential peaks 
are not significantly affected by $\GL$.  In both models, smaller values
of $\GL$ cause the membrane potential to take longer to return to $\Er$ after
an action potential (Fig. \ref{fig_HH_Cl__GL_range}); however, the post-peak 
oscillations
are larger for larger $\GL$ in the \Cl{} model, while they are smaller for
larger $\GL$ in the \Na/\K{} model.

We also investigated the effects of $\GL$ on the metabolic energy consumption
associated with action potentials, in view of other work \cite{crotty_2006b,levy_2006}
suggesting that the overall scale of the three conductances in the squid giant
axon is optimized for the energy associated with action potential velocity.
We did not find any such optimizations for $\GL$ alone, but only a monotonic
increase in metabolic energy consumption with increasing $\GL$.

\subsection{Maximum Firing Frequency}

\begin{figure}
\resizebox{3.25in}{2.5in}{\includegraphics{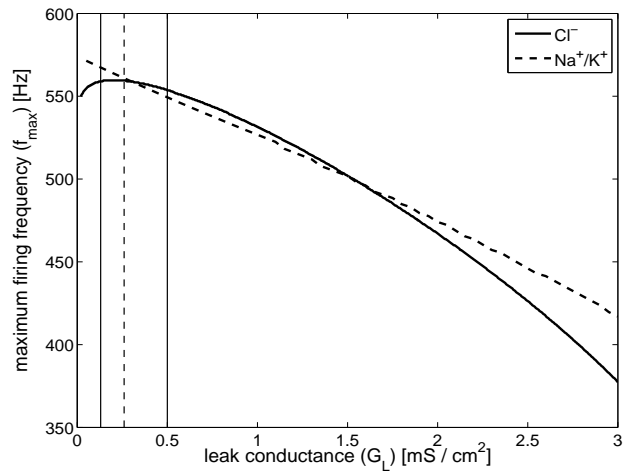}}
\caption{\label{fig_HH_Cl__GL_fmax} For \Cl{} leak channels, the 
maximum firing frequency has a maximum value of about 560 Hz when
$\GL$ is around 0.2 \mScmsq, in the range of experimentally measured values.  By
contrast, there is no local maximum for \Na/\K{} leak channels:  $f_{max}$ increases
monotonically with decreasing $\GL$.  Here and in subsequent figures, the two 
solid vertical lines show the range of $\GL$ measured by Hodgkin and Huxley, while 
the dashed vertical line shows the mean value of their measurements.  The $\GL$ 
resolution in all simulations is 0.005 \mScmsq.}
\end{figure}

\begin{figure}
\resizebox{3.25in}{2.5in}{\includegraphics{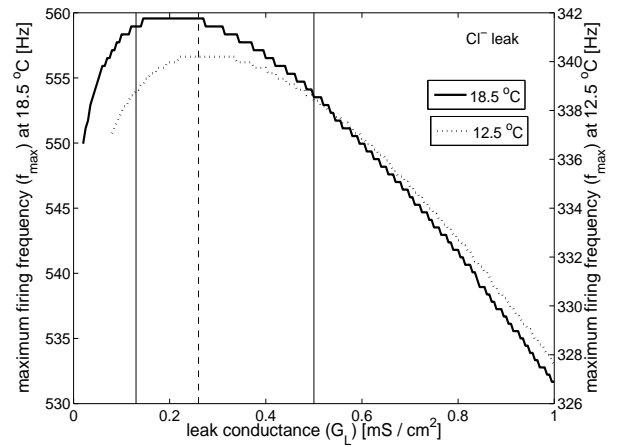}}
\caption{\label{fig_HH_Cl__GL_fmax_closeup} The region around the \Cl{} leak
model $f_{max}$ maximum in Figure \ref{fig_HH_Cl__GL_fmax} is shown in greater 
resolution at two different temperatures.  The 18.5 $^o$C case has its maximum at 
$\GL \approx 0.2 \pm 0.06$.  At 12.5 $^o$C, the maximum $f_{max}$ is at 
$\GL \approx 0.27 \pm 0.06$ and has a value of about 340 Hz.  The jagged shape of 
the curve and flatness near the minimum are due to the numerical limits of the 
simulation.}
\end{figure}

Figures \ref{fig_HH_Cl__GL_fmax} and \ref{fig_HH_Cl__GL_fmax_closeup} show
some of our central results, the maximum firing frequency, $f_{max}$, 
calculated as a function of $\GL$.  In the \Cl{} model, for values of $\GL$ much 
above the experimental range, $f_{max}$ decreases by about 80 Hz for each 
1 \mScmsq{} increase in $\GL$ (although the rate of decrease is slightly 
super-linear).  

However, as is just barely visible in Fig. \ref{fig_HH_Cl__GL_fmax} and 
much more evident in Fig. \ref{fig_HH_Cl__GL_fmax_closeup}, the relationship between 
$f_{max}$ and $\GL$ in the \Cl{} model is not monotonic.  For very low values of 
$\GL$, $f_{max}$ instead increases with $\GL$, attaining a maximum 
value of about 560 Hz at 18.5 $^\circ$C near $\GL = 0.2$ \mScmsq.  
Within the numerical limits of our simulation, the $f_{max}$-optimal value of $\GL$ 
for the \Cl{} model is about $0.2 \pm 0.06$ \mScmsq, well within the range of 
experimentally measured values.

In the \Na/\K{} model, however, $f_{max}$ decreases monotonically by about 60 Hz 
for every 1 \mScmsq{} increase in $\GL$.  This may be due to the different way 
$\GL$ affects the action potentials in this case (Fig. \ref{fig_HH_Cl__GL_range}).
The optimal value of $\GL$ in this model for maximum firing frequency is therefore
0.

Since the rate coefficients (Eqs. (\ref{eq_alpha_m}--\ref{eq_beta_n})) have a strong
dependence on temperature, we repeated our calculations at 12.5 and 25 $^\circ$C,
representative of the range of temperatures the {\it Loligo} squid genus studied
by Hodgkin and Huxley would normally experience.  At both of these temperatures 
and for both of the models, we found qualitatively similar 
behavior as at 18.5 $^\circ$C.  In the \Cl{} model, the maximum values of 
$f_{max}$ occur in or near the experimental range of $\GL$ 
(although the value of the maximum $f_{max}$ itself increases substantially with 
temperature, from about 340 Hz at 12.5 $^\circ$C to 848 Hz at 25 $^\circ$C).  The 
optimal value of $\GL$ decreases with increasing temperature from 
$\GL \approx 0.27 ~\textrm{\mScmsq}$ at $12.5 ~^\circ\textrm{C}$ 
(Fig. \ref{fig_HH_Cl__GL_fmax_closeup}) to 
$\GL \approx 0.11 ~\textrm{\mScmsq}$ at $25 ~^\circ\textrm{C}$.  In both cases,
the area around the maximum is fairly flat (although slightly less so for the
25 $^\circ$C case) with a width of about 0.06 \mScmsq.  While the optimal $\GL$ is
therefore not temperature-independent, the relative flatness of the maxima
means that the maximum firing frequency for values of $\GL$ near, 
for example, 0.2 \mScmsq{} are either at the maximum or within 2-3 Hz of it.

In the \Na/\K{} case, the relationship between $\GL$ and $f_{max}$ at the
other temperatures is still monotonically decreasing with no local maximum.
Additionally, no local maxima were observed when the magnitudes of the
active sodium and potassium conductances, rather than the leak conductance,
were varied.

\subsection{Repetitive Firing Frequency}

It is known \cite{rinzel_1978} from both theory and experiment that a constant
current input to one end of a non-space-clamped axon can produce repetitive firing 
at a constant frequency, albeit only over a fairly narrow range of current.
We investigated how the value of $\GL$ affects this firing frequency, $f_r$, which is 
qualitatively different than the firing produced by current spikes discussed above.

\begin{figure}
\resizebox{3.25in}{2.5in}{\includegraphics{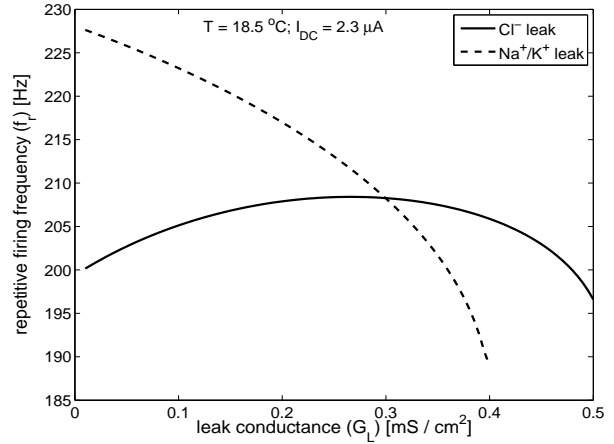}}
\caption{\label{fig_HHboth__GL_fr}  The repetitive firing frequency, $f_r$, 
i.e. the action potential frequency caused by a constant current input ($I_{DC}$) of
2.3 \uA{} at one end of the axon, is shown as a function of $\GL$.  As before, no 
local maximum is evident for the \Na/\K{} leak channel model, and for it $f_r$ 
increases with decreasing $\GL$.  However, the \Cl{} model once again attains a 
maximum well above vanishing $\GL$.  For this value of $I_{DC}$ and over a fairly 
large range in general, the maximum is very near the experimental value of $\GL$.} 
\end{figure}

For the \Na/\K{} channel model, as with the maximum firing frequency, 
we found no evidence of an optimization of $\GL$ at non-zero values for $f_r$.
The picture is considerably more complex, however, for the \Cl{} model.  Over
a fairly large range of input current and temperature, $f_r$ attains its
maximum value in or near the experimental range of $\GL$.  In Fig.
\ref{fig_HHboth__GL_fr}, we show $f_r$ versus $\GL$ for both models at
typical values of temperature and input current.  The \Cl{} $f_r$ maximum
in this case is about 208 Hz at $\GL$ = 0.265 \mScmsq.

\begin{figure}
\resizebox{3.25in}{2.5in}{\includegraphics{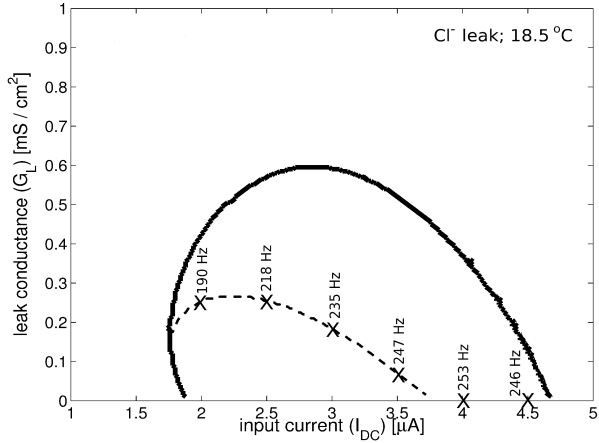}}
\caption{\label{fig_HH_Cl__IDC_GL_fr}  We show the behavior of the repetitive
firing frequency in two-dimensional ($\GL$-$I_{DC}$) parameter space.  The
solid black curve shows the boundary of the region outside which a constant
input current to one end of the axon does not produce sustained repetitive
firing.  The dashed curve inside the boundary shows the optimal value of $\GL$
for $f_r$ as a function of $I_{DC}$.  The numbers along the curve show the
actual value of $f_r$ at various points.  For the upper third of the $I_{DC}$ range
inside the boundary, the highest value of $f_r$ is attained when $\GL$ vanishes;
for most of the rest, however, the optimal $\GL$ is in the experimental range.}
\end{figure}

However, the value of $I_{DC}$ is a second independent parameter (assuming we hold 
all others fixed), and thus, in contrast to the maximum firing frequency produced
by discrete current spikes (which is independent of their actual size),  we must 
analyze the repetitive firing frequency as a function of both the leak 
conductance and the constant input current.  In Fig. \ref{fig_HH_Cl__IDC_GL_fr}, we 
show the repetitive firing frequency in the two-dimensional $\GL$-$I_{DC}$ parameter 
space.  The bold curve shows the limits of the region where repetitive firing can 
occur.  For combinations of $\GL$ and $I_{DC}$ outside it, repetitive firing either 
does not occur or lasts for only a few spikes.
The dashed curve within shows the $f_r$-optimal value of $\GL$ as a function of
$I_{DC}$; e.g., at $I_{DC}$ = 2.5 \uA, the maximum $f_r$ of 218 Hz occurs when
$\GL$ is about 0.255 \mScmsq.  If $\GL$ is larger or smaller than this, $f_r$
decreases in a way similar to what is shown in Fig. \ref{fig_HHboth__GL_fr}.

One evident feature of Fig. \ref{fig_HH_Cl__IDC_GL_fr} is that no repetitive firing 
at all is possible for values of $\GL$ above about 0.6 \mScmsq, regardless of
the value of $I_{DC}$.  This maximum upper limit of $\GL$ depends on 
temperature, as we will discuss, but is always below about 1 \mScmsq{} for
temperatures above about 10 $^\circ$C.  Thus, $\GL$ must be much smaller than
the maximum voltage-gated sodium and potassium conductances in order for 
repetitive firing to occur.

Another feature is that over about the lower half of the range of $I_{DC}$
where repetitive firing is possible, the optimal $\GL$ is within the
experimental range (0.13 to 0.5 \mScmsq), and is everywhere below
0.27 \mScmsq.  However, for values of $I_{DC}$ above about 3.8 \uA, the 
maximum $f_r$ obtains when $\GL$ is 0.  This includes the overall maximum $f_r$,
253 Hz, located at $I_{DC}$ = 4.08 \uA.  This value of $f_r$ is several tens of Hz 
above the maximum $f_r$ values at lower ranges where the optimal $\GL$ is in the
experimental range.

\begin{figure}
\resizebox{3.25in}{2.5in}{\includegraphics{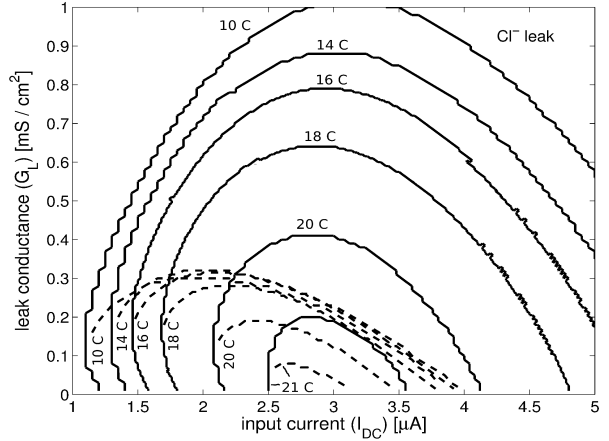}}
\caption{\label{fig_HH_Cl__T_range__IDC_GL_fr} We show the same information as 
in Figure \ref{fig_HH_Cl__IDC_GL_fr} but for a range of temperature characteristic
of what squid in the ocean encounter over periods of a few months.  In general,
higher temperatures decrease the size of the region of parameter space in which
repetitive firing occurs.  The optimal $\GL$ value for a given $I_{DC}$ is 
relatively temperature-independent except at high temperatures or low $I_{DC}$
values.}
\end{figure}

In Figure \ref{fig_HH_Cl__T_range__IDC_GL_fr}, we show the combined results for
an approximately 10 $^\circ$C range of temperature characteristic of what 
ocean-dwelling squid encounter over the course of a few months \cite{rosenthal_2000}.
At warmer temperatures, the range of $\GL$-$I_{DC}$ parameter space over which
repetitive firing is possible decreases sharply.  Conversely, the $f_r$-optimal
$\GL$ value as a function of $I_{DC}$ is relatively independent of temperature
below the highest temperatures or above the lowest $I_{DC}$ values.  Above about
4 \uA, the optimal $\GL$ is 0, with the highest absolute $f_r$ value also generally
in this range, while for most of the range below, $\GL$ is within the experimental 
limits.

\subsection{Relative Refractory Period}

\begin{figure}
\resizebox{3.25in}{2.5in}{\includegraphics{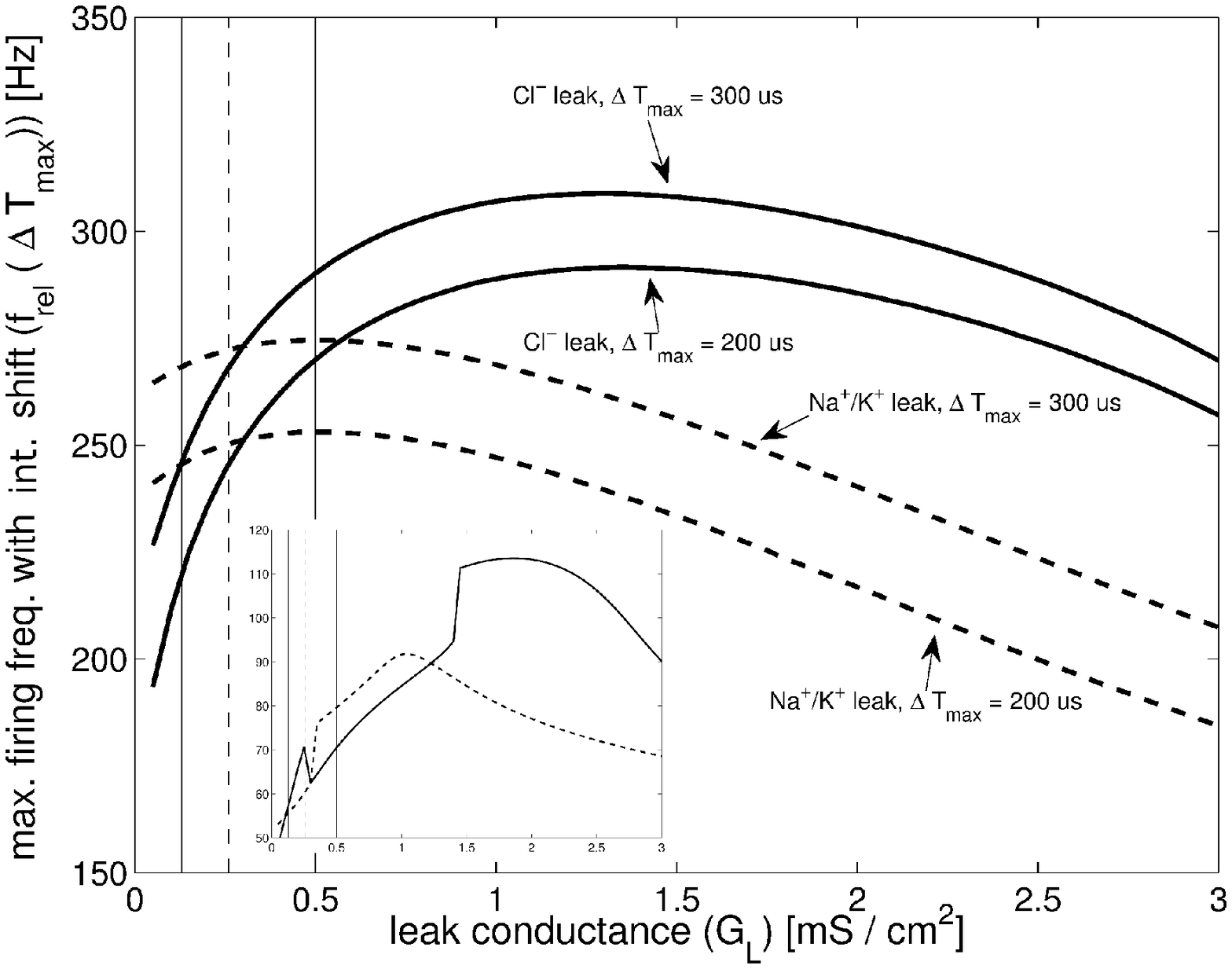}}
\caption{\label{fig_HH__x_0p8__dt_0p2__GL_frel} The maximum firing frequency
for a maximum allowed interval shift $\Delta T_{max}$ (defined in Eq. 
(\ref{eq_Trel_frel})) is shown
for the two leak channel models for $\Delta T_{max}$ = 200 and 300 \us.  For
\Cl{} channels, the frequency-optimal $\GL$ is in the range of 1.2-1.3 \mScmsq,
well above the experimental range.  For \Na{} and \K{} channels, the optimal $\GL$ is
lower but still at the upper end of the experimental range.  {\it Inset}:  we
show the frequencies for $\Delta T_{max}$ = 2 \us.  At this as well as at
other low values of the maximum interval shift, the frequency-optimal $\GL$ values 
are well outside the experimental range in both the \Cl{} (solid) and \Na/\K{}
(dashed) models.} 
\end{figure}

As discussed above, the relative refractory period is more easily and probably
more relevantly calculated with respect to a non-zero maximum interval shift
$\Delta T_{max}$.  In Fig. \ref{fig_HH__x_0p8__dt_0p2__GL_frel}, we show
the corresponding maximum firing frequency, $f_{rel} (\Delta T_{max})$, 
for a very low $\Delta T_{max}$, 2 \us{} (inset), and then for two higher
values, 200 and 300 \us.  The higher-$\Delta T_{max}$ curves have optimal
$\GL$ values substantially outside the experimental range in the case of
\Cl{} channels, and only barely inside it for \Na/\K{} channels.  For both
models, the values of these optima are insensitive to $\Delta T_{max}$ provided 
it is larger than about 150 \us.  The \Na/\K{} optima also appear to be relatively 
insensitive to other parameters such as temperature and axial resistivity.

\begin{figure}
\resizebox{3.25in}{2.5in}{\includegraphics{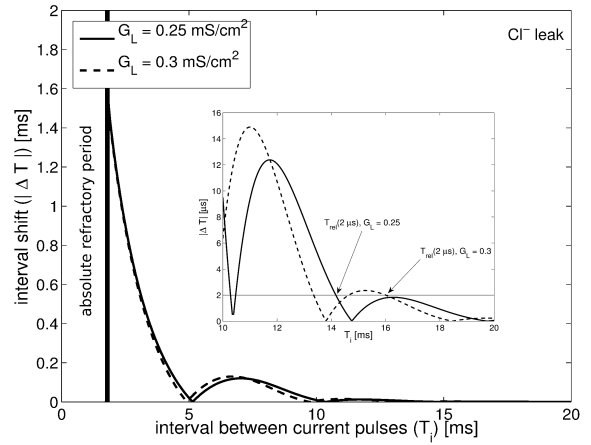}}
\caption{\label{fig_rrel_difficulties} The absolute value of the interval
shift, $|\Delta T| = |T_{AP} - T_i|$, is shown as a function of $T_i$ for two
nearby values of $\GL$.  The secondary peaks after $T_i$ = 5 ms are due to
the post-peak oscillations of the first action potential, which can either speed up
or slow down the second action potential.  {\it Inset}:  magnified view of
the $10 ~\textrm{ms} \le T_i \le 20 ~\textrm{ms}$ region.  The horizontal line
shows $|\Delta T| = \Delta T_{max} = 2 ~\textrm{\us}$; the largest $T_i$ value at 
which it intersects each curve gives $T_{rel}(\Delta T_{max})$.} 
\end{figure}

The low-$\Delta T_{max}$ curves look substantially different, and the reason
for this is illustrated in Figure \ref{fig_rrel_difficulties}.  The relative 
refractory period for a given $\Delta T_{max}$ is calculated by finding the largest 
value of the initial interval between current stimuli, $T_i$, such that 
$|\Delta T| = \Delta T_{max}$; we then define this $T_i$ value as 
$T_{rel}(\Delta T_{max})$, which can be visualized as the last point where the
horizontal line representing $\Delta T_{max}$ intersects the $T_i$ vs. $|\Delta T|$
curve.  For values of $\Delta T_{max}$ larger than about 150 \us, this point always 
falls on the first, monotonically decreasing part of the curve below 
$T_i \approx 5 ~\textrm{ms}$, resulting in the smooth appearance of the 
high-$\Delta T_{max}$ curves in Fig. \ref{fig_HH__x_0p8__dt_0p2__GL_frel}.  For low 
values of $\Delta T_{max}$, however, the point of intersection falls within the
region of secondary peaks due to the membrane potential oscillations after the 
first action potential.  The sizes and locations of the secondary peaks
depend on $\GL$.  As a result, it is possible for a peak to be just above
$\Delta T_{max}$ for one value of $\GL$ and just below it for a second, nearby value
of $\GL$, causing a sharp transition in the value of $T_{rel}(\Delta T_{max})$ as
a function of $\GL$.  This is the case in the inset to Fig. 
\ref{fig_rrel_difficulties}:  for $\GL$ = 0.3 \mScmsq, the point of intersection
is at about $T_i$ = 16 ms, while for $\GL$ = 0.25 \mScmsq, the size of the peak there
is slightly below $\Delta T_{max}$ = 2 \us{} and therefore the intersection point
jumps down to the previous oscillation, at about $T_i$ = 14.2 ms.  We therefore
see a sharp spike at $\GL$ = 0.25 \mScmsq{} in the inset to 
Fig. \ref{fig_HH__x_0p8__dt_0p2__GL_frel} for \Cl{}.  Similar effects are
seen in any curve of $f_{rel} (\Delta T_{max})$ vs. $\GL$ for small values of
$\Delta T_{max}$.   

\section{Discussion}

The two scenarios considered here for the stimulation of action potentials in
the squid giant axon, delta function-like current spike inputs and 
unchanging constant current inputs, represent idealized extremes of the actual 
biology.  The input current to the squid giant axon originates from postsynaptic 
glutamate-activated sodium channels at the squid giant synapse \cite{gage_1969}.  
The frequency and duration with which these currents are evoked are ultimately 
determined by the squid's sensory environment, e.g., whether it perceives any 
predators to be nearby.  Thus, the actual current input to the axon when it is active 
is neither instantaneous nor constant, but is likely to be a time-varying function 
determined by the rate of synaptic bombardment and the kinetics of the synapse and 
postsynaptic sodium channels.  Because of the relative dearth of experimental data on 
the operation of the squid giant axon system {\it in vivo} \cite{preuss_2000},
it is difficult to accurately model this current.  However, it might
qualitatively be expected that optimization results which hold for the two
theoretical extremes would hold for the true time-dependent input current
to the axon as well. 

Our results indicate that if the input stimulus to the axon can be regarded as
a series of discrete, sharp pulses, and if the leak current is assumed to be
chloride, then the maximum firing frequency of the axon is itself maximized for 
values of leak conductance within the experimental range of values.  For continuous
input currents ($I_{DC}$), the same is generally true for at least half of the range 
of $I_{DC}$ and almost all the range of temperature over which repetitive firing can 
occur.  However, when one considers the full range of $\GL$-$I_{DC}$ parameter space, 
the repetitive firing frequency is maximized overall at high values of 
$I_{DC}$ and $\GL$ = 0.  This raises the question of why, if the input current is 
constant (or, at least, has a timescale much longer than the repetitive firing 
frequency), a vanishing leak conductance and high input current level would not be 
preferable.

One possible reason is metabolic energy consumption:  higher $I_{DC}$ values would be 
associated with larger currents through the \Na{} channels in the 
synapse as well as more frequent action potentials in the axon itself.  This implies 
that the metabolic energy cost of driving the axon at a frequency $f_r$, which can be 
quantified by the overall flux of \Na{} ions into the membrane (all of which has to 
be subsequently pumped back out by the ATPase \Na/\K{} exchanger in order to restore 
the resting concentration gradient), would be substantially higher at higher $I_{DC}$ 
and $f_r$ values.  Of course, presumably it would not be optimal for $f_r$ itself to 
be too low.  Hence, there may be an optimization involving both firing frequency
and metabolic energy which would favor the lower end of the $I_{DC}$ range and
the observed values of $\GL$.

A simpler and perhaps more compelling reason is the requirement that the
squid giant axon be able to function over a range of ocean temperature that can
span 10 or more degrees Celsius over the course of a few months \cite{rosenthal_2000},
and several degrees over a single day \cite{kawai_2007}.  An important consideration
is that both $\GL$, which depends on the leak channel density on the axon, and 
$I_{DC}$, which depends on the amount of neurotransmitter released per 
presynaptic action potential and the density of receptors on the 
postsynaptic membrane, are unlikely to be quickly changeable in response to
a temperature change.  (Even if the input current is not constant in time,
its maximum or typical amplitude, of which we are taking $I_{DC}$ as a rough
estimate, would depend on these properties and not be quickly changeable.)
That is, for our purposes, we will assume that both $\GL$ and $I_{DC}$ are 
essentially fixed.

With these assumptions, Fig. \ref{fig_HH_Cl__T_range__IDC_GL_fr} makes it
clear that values of $I_{DC}$ in the vicinity of 3 \uA{} are preferable to values
much lower or higher in order for repetitive firing to be possible over the widest
possible temperature range.  For example, if $I_{DC}$ were 4 \uA{} and $\GL$ 0, which 
gives the maximum repetitive firing frequency at 18.5 $^\circ$C, a modest temperature
increase of only 1.5 $^\circ$C would put $I_{DC}$ and $\GL$ outside the region
of parameter space where repetitive firing can occur.  The axon would be 
rendered inoperable.

Assuming that $I_{DC}$ must be around 3 \uA{} in order for the squid giant axon to
function effectively at warmer temperatures means $\GL$ should be at a value
which in general gives the highest possible $f_r$ value at that $I_{DC}$.  We
can see from Fig. \ref{fig_HH_Cl__T_range__IDC_GL_fr} that over most of
the temperature range, the optimal $\GL$ values at $I_{DC} \approx$ 3 \uA{} are
clustered around 0.2 \mScmsq.  Therefore, we may expect that $\GL$ values in
the experimental range are most optimal for repetitive firing given the typical 
temperature variations the squid encounters.  In sum, for \Cl{} leak channels, 
biological values of $\GL$ appear optimal for maximizing the maximum or repetitive 
firing frequency of the axon as determined by the Hodgkin-Huxley model.

The same is not true, however, of \Na/\K{} leak channels, which show no such 
optimization at non-zero $\GL$ values for the maximum or repetitive firing 
frequencies.  They only appear to be superior in this regard to \Cl{} channels 
when it comes to the relative refractory period as defined for a maximum
allowed interval shift above about 150 \us:  while both models evince non-zero 
optima for $\GL$, the one for \Cl{} is far above the biological range of values, 
and the one for \Na/\K{} is only just within it.  However, the calculation
of relative refractory periods is problematic for lower values of the maximum
interval shift.  Moreover, the relevance of the relative refractory period,
which characterizes the maximum firing frequency without information loss,
to a peripheral axon like the squid giant axon is not clear.  Thus, we consider
our results for the firing frequency associated with the relative refractory 
period to be much less compelling than those for the absolute and repetitive 
firing frequencies.

\section{Conclusions}

If one assumes that it is best for the squid's brain to be able to send two or 
more signals to its escape jet system with as little delay between them as
possible, and also that the Hodgkin-Huxley model is a sufficiently accurate
model of the biological squid giant axon, then the experimentally measured range of 
values for the squid giant axon leak conductance make far more sense for a 
chloride-like leak current than for one composed of separate sodium and potassium 
leak currents.  The leak conductance appears to be optimal for for the firing rate 
of the axon, whether it be driven by discrete input current pulses or a by constant 
input current, if the leak current is assumed to be chloride or some combination 
of ions whose overall reversal potential is approximately $\ECl$.  If the leak 
current is instead assumed to be composed of separate sodium and potassium currents, 
then no such optimization is evident, though there is weak evidence of a partial 
optimization for relative refractory period.  It should be remembered, though, that 
these results are all within the context of the Hodgkin-Huxley model.

Of considerably more interest than the evolutionary neurobiology of {\it Loligo}
is whether such an optimization of the leak conductance for firing rates exists or 
has taken place in mammalian central neurons.  Due to the current lack of precise 
data on channel densities and kinetics in these much smaller and more morphologically 
complex cells, it is hard to address such questions in a rigorous way with
modeling studies.  Nevertheless, whenever such data become available, it may be
fruitful to examine in detail the role of the leak conductance on firing frequencies
and information rates, as it appears to be a powerful mechanism for influencing
these properties despite its deceptively small magnitude.


\begin{acknowledgments}
We thank J. Amato, K. Andresen, K. Belanger, W. B. Levy, J. Meyers, and 
K. Segall for useful discussions, as well as our referees at Physical Review
E for reviewing our paper.
\end{acknowledgments}


\end{document}